\documentclass[12pt]{amsart}
\usepackage{mathrsfs}
\usepackage{amssymb} 
\usepackage{url}
\usepackage{cite}
\usepackage{hyperref}
\usepackage{color}
\usepackage{hyperref}
\definecolor{refcolor}{RGB}{0,0,190}
\hypersetup{
    colorlinks,
    citecolor=refcolor,
    filecolor=refcolor,
    linkcolor=refcolor,
    urlcolor=refcolor
}

\def\({\left(}
\def\){\right)}

\newcommand{\op}[1]{\operatorname{#1}}

\newcommand{\de}{\op{d}}

\newcommand{\tn}{\textnormal}
\newcommand{\ds}{\displaystyle}

\newcommand{\dsfrac}[2]{\ds{\frac{#1}{#2}}}

\makeatletter
\renewcommand\section{\@startsection {section}{1}{\z@}%
                                   {-3.5ex \@plus -1ex \@minus -.2ex}%
                                   {2.3ex \@plus.2ex}%
                                   {\center\normalfont\Large\bfseries}}
\renewcommand\subsection{\@startsection {subsection}{2}{\z@}%
                                   {-3.5ex \@plus -1ex \@minus -.2ex}%
                                   {2.3ex \@plus.2ex}%
                                   {\normalfont\large\bfseries}}
\renewcommand\subsubsection{\@startsection {subsubsection}{3}{\z@}%
                                   {-3.5ex \@plus -1ex \@minus -.2ex}%
                                   {2.3ex \@plus.2ex}%
                                   {\normalfont\bfseries}}
\makeatother

\begin{document}

\title{Kaluza theory with zero-length extra dimensions}

\author{O.C. Stoica*}
\address{Department of Theoretical Physics, \\National Institute of Physics and Nuclear Engineering -- Horia Hulubei, Bucharest, Romania.}
\email{cristi.stoica@theory.nipne.ro; holotronix@gmail.com}

\date{\today \\ *Department of Theoretical Physics, National Institute of Physics and Nuclear Engineering -- Horia Hulubei, Bucharest, Romania. E-mail address: cristi.stoica@theory.nipne.ro; holotronix@gmail.com}
\maketitle

\begin{abstract}
A new approach to the Kaluza theory and its relation to the gauge theory is presented. Two degenerate metrics on the $4+d$-dimensional total manifold are used, one corresponding to the spacetime metric and giving the fiber of the gauge bundle, and the other one to the metric of the fiber and giving the horizontal bundle of the connection. When combined, the two metrics give the Kaluza metric and its generalization to the non-abelian case, justifying thus his choice. Considering the two metrics as fundamental rather than the Kaluza metric explains why Kaluza's theory should not be regarded as five-dimensional vacuum gravity. This approach suggests that the only evidence of extra dimensions is given by the existence of the gauge forces, explaining thus why other kinds of evidence are not available. In addition, because the degenerate metric corresponding to the spacetime metric vanishes along the extra dimensions, the lengths in the extra dimensions is zero, preventing us to directly probe them. Therefore this approach suggests that it is not justified to search for experimental evidence of the extra dimensions as if they are merely extra spacetime dimensions. On the other hand the new approach uses a very general formalism, which can be applied to known and new generalizations of the Kaluza theory aiming to achieve more and make different experimental predictions.
\end{abstract}

\section{Introduction}
\label{s_intro}

The {\em Kaluza theory} is a unification of general relativity and electrodynamics, based on an idea proposed by Nordstr{\"o}m in his metric theory of gravity \cite{nordstrom1914moglichkeit}.
The general relativistic version was proposed by Kaluza \cite{kal:1921}.
At that time it was considered that the fifth dimension should be observable, which would contradict our experience.
To explain why the fifth dimension is not observed, and also to unify charge and momentum, Klein proposed that it was {\em compact and very small} \cite{kle:1926}.

The literature on the Kaluza-Klein theory is rich and the research is still very active, the theory being at the origin of new directions like supergravity and superstring theory. A recent review of three main directions, namely compactified, projective, and noncompactified Kaluza-Klein theories, can be found in \cite{OverduinWesson1997KaluzaKleinGravity}~.

After the geometric formulation of gauge theory in terms of principal bundles, it was realized that there must exist a relation between the two theories.
Here we will present a new way to derive Kaluza's metric from gauge theory. The proposed approach applies to any gauge groups, including non-abelian ones, so it also works for the gauge group of the standard model of particle physics. We will derive the generalization of the Kaluza theory proposed in \cite{Kerner1968KaluzaKleinGaugeGroup}.

The method which we will employ here relies on using of two degenerate metrics on the $4+d$-dimensional total manifold, one  corresponding to the spacetime metric and whose radical is the fiber, and the other one to the metric of the fiber, and whose radical is the horizontal space of the connection.

In the following we will briefly recall some known results on degenerate metrics. Then, we will apply such metrics to obtain the Kaluza metric. We will discuss the implications on the possible reason for lacking experimental evidence for the extra dimensions, and also the possible generalizations.

A \emph{degenerate} metric is a symmetric bilinear form which is degenerate.
Spaces with degenerate metrics were studied in \cite{Barb39,Moi40,Str41,Str42a,Str42b,Str45,Vra42}.

Until recently, the state of the art was the work of D. Kupeli \cite{Kup87b,Kup96}.
But there were two limitations:
\begin{itemize}
	\item the signature was constant, while in general relativity has to change at singularities,
	\item the theory was not invariant, and it depended on the choice of a distribution transversal to $\ker g$.
\end{itemize}
The results from
\cite{Sto11a}
apply to changing signature, are invariant, and don't rely on a particular choice. The particular cases of Kupeli and Riemann are obtained. However, for the purpose of this article it is enough to work with constant signature.

\section{Gauge theory and Kaluza theory}

Let $(E,M,\pi,F)$ be a {\em fiber bundle} with {\em total space} $E$, {\em fiber} $F$, {\em base space} $M$, and projection $\pi:E\to M$.

If $M$ is a semi-Riemannian manifold with metric $g$, then $g$ defines on the total space $E$ uniquely a degenerate metric $\widetilde g$, as the pull-back of $g$,
$$\widetilde g=\pi^\ast g.$$

Let $V<TE$ be the {\em vertical bundle}, $V:=\ker(\de \pi)$.
Then, at every point $p\in E$, the vertical tangent space $V_p$ is the radical of $\widetilde g_p$. So we have
$$\ker(\de\pi)=\ker{\widetilde g}.$$

Let now $(E,M,\pi,F,G)$ be a {\em principal $G$-bundle}, where the typical fiber $F$ is a {\em $G$-torsor} (hence is diffeomorphic to $G$ and $G$ acts freely and transitively on $F$).

For gauge theory, the principal bundle has to be endowed with a horizontal distribution $H<TE$ defining a {\em gauge connection}, and with a metric $\widetilde h_V$ on the tangent bundle of the typical fiber, both gauge invariant with respect to the transformations from $G$. The two objects $H$ and $\widetilde h_V$ are equivalent to a metric $\widetilde h$ on $TE$ which is degenerate on $H$ and gauge invariant, defined at any point $p\in E$ by
\begin{equation}
\widetilde h(X,Y) = \widetilde h_V(\pi_V X,\pi_V Y),
\end{equation}
where $\pi_V:TE\to V$ is the projection on $V$ along $H$. Then, 
$$H=\ker{\widetilde h}.$$

The restriction of the metric $\widetilde g$ on the horizontal distribution $H$ is a non-degenerate metric $\widetilde g_H$.
Since $\widetilde g|_V=0$, from $H,V$ and $\widetilde g_H$ one can recover $\widetilde g$ by
$$\widetilde g(X,Y) = \widetilde g_H(\pi_H X,\pi_H Y),$$
where $\pi_H:TE\to H$ is the projection on $H$ along $V$.

The Kaluza theory can be seen now as combining the two metrics $\widetilde g_H$ on $H$ and $\widetilde h_V$ on $V$ into a metric on $E$ given by
\begin{equation}
\hat g_0(X,Y)=\widetilde g(X,Y) + \widetilde h(X,Y),
\end{equation}
whose components in a frame composed of a horizontal and a vertical frame has the form
\begin{equation}
\label{eq:kaluza_metric_HV}
\hat g_0 =	\left(
\begin{array}{cc}
	g_{ab} & 0 \\
	0 & h_{\alpha\beta} \\
\end{array}
\right).
\end{equation}

We can identify $E$ at least locally to the product $E=M \times F$. Then to obtain the metric $\hat g_0$ on $M \times F$ we apply a transformation that leaves the fibers invariant, and projects the horizontal space $H_p$ to the space $T_pM$,

\begin{equation}
S =	\left(
\begin{array}{cc}
	I_4 & A \\
	0 & I_d \\
\end{array}
\right),
\end{equation}
where $A=A_a^\mu$ is the connection determined by $H$, and $d=\dim G$.

Then, the metric \eqref{eq:kaluza_metric_HV} in a frame compatible to the product $M\times F$, for example which is the product of a frame of $M$ and one of $F$, has the form
\begin{equation}
\label{eq:kaluza_metric}
	\hat g_{ij} = S \hat g_0 S^T =
	\left(
\begin{array}{cc}
	g_{ab} + h_{\mu\nu} A_a^\mu A_b^\nu & h_{\mu\beta} A_a^\mu \\
	h_{\alpha\nu} A_b^\nu &  h_{\alpha\beta} \\
\end{array}
\right),
\end{equation}
where $g_{ab}$ is the Lorentzian metric on $M$.

We recover thus the generalized Kaluza theory for an arbitrary non-abelian gauge group (see Kerner \cite{Kerner1968KaluzaKleinGaugeGroup}~, for non-abelian Kaluza-Klein theory also see \cite{Trautman1970NonAbelianKaluzaKlein,deWitt1966NonAbelianKaluzaKlein,Cho1975HigherDimGauge}~, also compare to the derivation in terms of Nambu-Goldstone fields in \cite{Cho1975NonAbelianGauge}).

In particular, to obtain the original Kaluza theory, which unifies gravity and electromagnetism, one sets $G=U(1)$ and $h=1$:

\begin{equation}
\label{eq:kaluza_metric_em}
	\hat g =
	\left(
\begin{array}{cc}
	g_{ab} + A_a A_b & A_a  \\
	A_b  &  1 \\
\end{array}
\right).
\end{equation}

By taking as Lagrangian density the scalar curvature corresponding to $\hat g$,
one obtains the {\em Einstein-Maxwell equations}, that is,
the source-free Maxwell equations,
and the Einstein equation for the four-dimensional metric $g_{ab}$ with the stress-energy tensor 
\begin{equation}
\label{eq_stress_energy_maxwell}
	T_{ab} = \dsfrac{1}{\mu_0}\left(F_{as}F_{b}{}^s - \dsfrac 1 4 F_{st} F^{st}g_{ab}\right)
\end{equation}
sourced by the electromagnetic field.

\section{Interpretation of the proposed construction}

Our proposal distinguishes the extra dimensions as those along which the $4+d$ dimensional metric $\widetilde g$ is degenerate. It also identifies the horizontal bundle to the distribution on which the metric $\widetilde h$ is degenerate.
This approach leads immediately and naturally to the metric postulated by Kaluza. In the same time, the metric $\widetilde g$ gives naturally the projection in the fiber bundle formulation of gauge theory, and $\widetilde h$ gives the gauge connection.

If the two metrics $\widetilde g$ and $\widetilde h$ are considered to be the fundamental objects, the non-degenerate metric \eqref{eq:kaluza_metric} in which they can be combined should not be seen as fundamental. The reason is that at each point $p\in E$ the transformations preserving the two degenerate metrics $\widetilde g$ and $\widetilde h$ are different from those preserving the combined metric. Not only the Kaluza metric has to be preserved, but also the splitting of the tangent bundle of the $4+d$-dimensional space into horizontal and vertical bundles. The approach based on the metrics $\widetilde g$ and $\widetilde h$ leads to invariance with respect to the spacetime diffeomorphisms combined with the gauge transformations. While the equations are invariant also to general $4+d$-dimensional diffeomorphisms, these obfuscate the fiber bundle structure and the connection.

This explains why the Hilbert-Einstein action in $4+d$ dimensions has to be varied such that it preserves the horizontal and vertical bundles in order to obtain the correct Einstein-Maxwell equations. This should be contrasted with Kaluza's original idea that gravity and electromagnetism are unified as pure five-dimensional gravity, therefore postulating the vanishing of the five-dimensional Ricci tensor. If the Ricci tensor's component in the fifth dimension vanishes, this result in the equation $F^{st} F_{st}=0$, which means that the Lagrangian of the electromagnetic field vanishes, and at every point in spacetime the magnitudes of the electric and magnetic vector fields are equal, which is too restrictive. This problem is well known and there are other ways to fix it (see for example \cite{OverduinWesson1997KaluzaKleinGravity}~ \S 3.4 and \cite{Wehus2004ScalarKaluzaKlein}).

The metric $\widetilde g$ corresponding to spacetime vanishes along the extra dimensions, therefore the size of these dimensions is zero. So this approach predicts that any attempt to probe the extra dimensions by measuring their size or distances in the fibers fails. The only evidence of the extra dimensions comes from the metric $\widetilde h$, which gives the connection and the metric of the fiber. But this metric is different from the spacetime metric, so one should not expect to be possible to probe extra dimensions as if they are extra spacetime dimensions.
Consequently, the only possible evidence of extra dimensions in this theory is the existence of the gauge fields. When translated to the Kaluza theory, this results in the cylinder condition introduced by Kaluza, which says that all physical fields are independent of the additional dimensions. In conclusion, the gauge fields are evidence of the extra dimensions, but gauge symmetry also results in the cylinder condition, which prevents us from probing more than the gauge fields. Even by removing the cylinder condition, the metric $\widetilde g$ corresponding to spacetime, being degenerate along the extra dimensions, gives zero lengths in these dimensions. Therefore, even though up to this date there is no evidence of extra dimensions, our proposal is still not refuted.
In particular, because the metric $\widetilde g$ does not distinguish between points in the same fiber, it makes no sense to try to find predictions based on geodesic motion in the extra dimensions.
However, this does not preclude the possibility that generalized versions of it make different predictions, especially if we consider the Kaluza metric \eqref{eq:kaluza_metric} as being more fundamental than $\widetilde g$ and $\widetilde h$. Such predictions have to be within the current experimental constraints like those analyzed for example in \cite{Casalbuoni1999SMKaluzaKleinElectroweak,LiuOverduin2000SolarSystemTestsKaluzaKlein,CheungLandsberg2002KaluzaKleinGaugeBosons,Overduin2013ConstraintsKaluzaKleinGravityProbeB,Wesson2015ExtraDimensionsPhaseExeriments}~.

Generalizations in many directions are possible in the approach introduced here.
The highest generality is obtained by imposing a single condition to the degenerate metrics $\widetilde g$ and $\widetilde h$ -- that the tangent space of the $4+d$-dimensional manifold is the direct sum of their radicals, and give up the other conditions corresponding to gauge theory.
Therefore, the approach presented here extends in a simple way in the usual directions in which the Kaluza theory was generalized, for instance it allows easily the elimination of the cylinder condition and the compactification condition. The present results and the possibilities of applications to generalizations of the Kaluza theory justify future development of these ideas.

\section{Extensions of Kaluza-Klein theory and experimental constraints}

Another reason not to impose the Ricci flatness of the higher-dimensional space in order to get the Einstein-Maxwell or Einstein-Yang-Mills equations is that this would not work if sources have to be included. For instance, in the $\tn{U}(1)$ case, the electromagnetic current corresponds to the components of the Ricci tensor of the form $R_{a5}$.
Therefore, the Kaluza theory is incomplete, because the gauge force in this theory is sourceless. The source can be any charged particle (in the electromagnetic case), including Dirac electrons, and we do not know a good way to extend the source-free Kaluza theory to include them. Klein's method to include them is not satisfactory, because it leads to a too big mass for the electron. A more general theory than Kaluza's is desirable, and some proposals are discussed for example in \cite{Collins1989ParticlePhysicsAndCosmology,OverduinWesson1997KaluzaKleinGravity}~.
Recent extensions of the Kaluza-Klein theory aim to obtain more realistic and more complete descriptions, and to resolve some problems of the older Kaluza-Klein theories, as well as other problems in physics.
While the Kaluza theory is sourceless, it is possible to add more dimensions in order to describe charged fields which are the sources of the electromagnetic field in the Kaluza theory, or to introduce higher-derivative terms in the Hilbert-Einstein action (\cite{OverduinWesson1997KaluzaKleinGravity}~\S 5.2). A promising alternative is provided by projective Kaluza-Klein theory \cite{Schmutzer1983ProjectiveKaluzaKlein,Schmutzer1995ProjectiveKaluzaKlein,Schmutzer2009ReportProjectiveKaluzaKlein}~.
Some generalization of Kaluza's theory rely on introducing a scalar factor which multiplies the metric $h_{ab}$ in equation \eqref{eq:kaluza_metric}, or on relaxing the cylinder condition, or on removing the compactification of the fifth dimension. Notable in this respect is the {\em space-time-matter} theory \cite{Wesson2007SpaceTimeMatter}~, which aims to describe matter in a five-dimensional generalization of the Kaluza-Klein theory.
Another remarkable approach is based on the vertical and horizontal distributions, and a sub-Riemannian metric defined on the horizontal distribution, and can be applied to both Kaluza-Klein theory, and to space-time-matter theory \cite{Bejancu2012CurvatureSubRiemannian,Bejancu2013SpaceTimeMatter}~. Very interesting results, based on of connections and generalized Lie algebroids and connections, were obtained in \cite{arcucs2014pseudoKaluzaKlein}~.
Other developments are in the direction of superstring theory, which is believed by many to include the Standard Model, see for example the review article \cite{Raby2011StandardModelString}~.
Some models rely on branes on which the Standard Model particles are confined, while gravitons move freely in the extra dimension, in order to explain why gravity is so much weaker than the other forces (the {\em hierarchy problem}), including the prediction of extra dimensions of the order of one millimeter \cite{ADD1998ExtraDimensionOneMillimeter,ADD1999PhenomenologyExtraDimensionOneMillimeter,RandallSundrum1999}~.
These are only samples of a very wide and deep literature on the problem.
The predictions of many such models include ways to probe the extra dimensions, even if the needed energies will still be inaccessible to us for long time.

Various extensions of the Kaluza-Klein theory produce testable predictions, which can be constrained by the current observations.
For example, classical tests of general relativity were applied to Kaluza-Klein gravity, using an analog of the four-dimensional Schwarzschild metric, at the level of the solar system \cite{LiuOverduin2000SolarSystemTestsKaluzaKlein}~.
For a spinning test body in Kaluza-Klein gravity with one additional space dimension, constraints are obtained from Gravity Probe B \cite{Overduin2013ConstraintsKaluzaKleinGravityProbeB}~.
An experimental test in terms of electroweak data was considered, in order to restrict the parameters of a minimal extension to extra dimensions of the Standard Model, in \cite{Casalbuoni1999SMKaluzaKleinElectroweak}~.
Also in \cite{CheungLandsberg2002KaluzaKleinGaugeBosons}~ it is shown that data from the LEP 2, Tevatron, and HERA experiments, sets a lower limit on the extra dimension compactification scale to $> 6.8$ TeV at the $95\%$ confidence level, for models with the Standard Model gauge bosons propagating in TeV${}^{-1}$-size extra dimensions, whose Kaluza-Klein states interact with the rest of the SM particles
confined to the 3-brane.
Also, a proposal to obtain constraints for the null path approach by measurements of the Aharonov-Bohm effect using neutron interferometry was made in \cite{Wesson2015ExtraDimensionsPhaseExeriments}~. These are only a few examples of ways to test the various extra dimensions theories or to impose constraints on their parameters.

While up to this point relevant constraints could be obtained from the experimental data, there is still no direct evidence of extra dimensions or of their absence. This situation is compatible with many of these approaches which have the parameters within the observed limits, in particular with the one proposed here, which simply predicts that the only evidence of extra dimensions is gauge symmetry and gauge forces. Also, even if we remove the cylinder condition in our model, since the metric $\widetilde g$ is degenerate along the extra dimensions, the lengths as measured with $\widetilde g$ still vanish in the extra dimensions. 
If we will find evidence of positive size extra dimensions, this would falsify the hypothesis that the two degenerate metrics are fundamental.
But even in this case, the formalism developed here would still be valid, under the hypothesis that the Kaluza metric, which allows positive sizes of the extra dimensions, is the fundamental one.

\subsection*{Acknowledgement}

The author thanks a referee for valuable recommendations which improved the clarity and quality of the article.


\end{document}